# Effect of spatial constraints on fragment production

Pallavi Gupta*, Rajni and Suneel Kumar
*School of Physics and Material Science, Thapar University, Patiala 147004, Punjab, INDIA*
*email: guptapallvi@gmail.com

## Introduction

The study of heavy ion collision at intermediate energies is the tool to determine the bulk properties of nuclear matter, or nuclear equation of state, and the understanding of collision process, which vary over the intermediate energy range. This study gives us the possibility to understand the properties of nuclear matter at extreme condition of temperature and density.

Multifragmentation is the most extensively studied phenomena in this energy domain. In the past people have tried to develop various methods of clusterization[1]. Among these methods minimum spanning tree[1] is one of the fastest method. In the present study our aim is to understand the role of spatial constraints on fragmentation.

In this paper, we take range of clusterization parameter $R_{clus}$ is between 2 fm and 6 fm. By varying the value of $R_{clus}$ we see its effect on the multiplicity of fragments. After optimizing the value of $R_{clus}$, we have tried to compare the theoretical calculations with experimental data of ALADiN collaboration[2].

## IQMD model

For simulation we use IQMD model[3]. In this model the propagation is governed by the classical equation of motion:

$$\dot{r} = \frac{\partial H}{\partial p_i}; \qquad \dot{p} = -\frac{\partial H}{\partial r_i}$$

Where H stands for the Hamiltonian which is given by:

$$H = \sum_{ij}^{A} \frac{p_i^2}{2m} + \sum_{ij}^{A}(V_{ij}^{Sk} + V_{ij}^{Yu} + V_{ij}^{Cou} + V_{ij}^{mdi} + V_{ij}^{sym})$$

The $V_{ij}^{Sk}$, $V_{ij}^{Yu}$, $V_{ij}^{Cou}$, $V_{ij}^{mdi}$, and $V_{ij}^{sym}$ are, respectively, the Skyrme, Yukawa, Coulomb, momentum dependent interaction (MDI), and symmetry potentials.

## Results and discussion:

In the present analysis, we have simulated the symmetric reaction of $^{197}Au_{79} + ^{197}Au_{79}$, $^{131}Xe_{54} + ^{131}Xe_{54}$, $^{20}Ne_{10} + ^{20}Ne_{10}$ at incident energy of 100MeV/nucleon for central and semi central impact parameters. A soft and soft momentum dependent (SMD) equation of state is being used in these simulations. The reaction dynamics is followed until 200 fm/c and then clusterization is performed with MST method in which nucleon are bounded if $R_{clus} = |\vec{r}_i - \vec{r}_j| \leq d_{min}$. $d_{min}$ have been varied between 2 and 6 fm respectively.

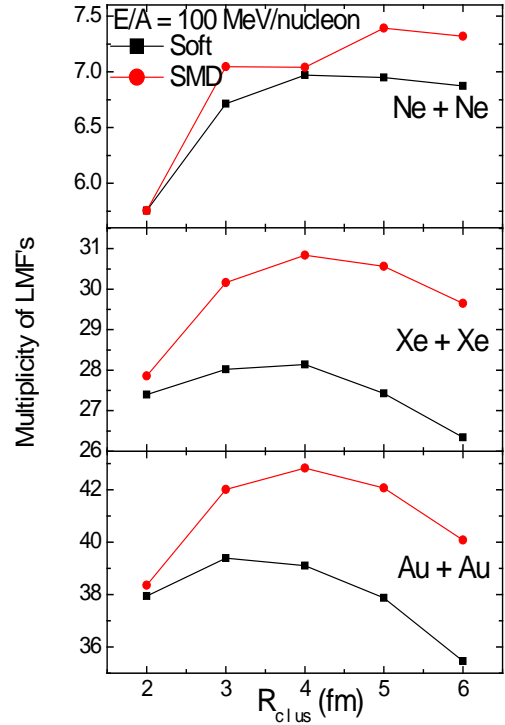

Fig.1. Variation of multiplicity of LMF's as function of $R_{clus}$ (fm) for central collision.

In fig.1, we check the multiplicity of light mass fragments (LMF's) for different value of $R_{clus}$ for soft and SMD equation of state. As we decrease the distance among the nucleon more free nucleons are produced and hence decrease in the production of light mass fragments (LMF's). For larger value of $R_{clus}$ more number of heavy mass fragments are produced. Thus in both the cases less production of LMF's.

For $R_{clus}$= 4 fm we obtain the maximum number of LMF's for all the colliding nuclei. Also the production of LMF's depend upon the size of colliding nuclei, for large nuclei, more number of LMF's are produced as compared to small nuclei. From the figure it is clear that enhanced production of LMF's take place in the presence of momentum dependent interaction as compare to static one. This is due to repulsive nature of the MDI's, which help in breaking the heavier fragments into LMF's.

observed. For central geometry the collisions are violent so there are few number of IMF's are observed due to disassembly of nuclear matter.

At peripheral collisions very small portion of target and projectile overlap so again few number of IMF's observed most of the fragments goes out in the form of heavy mass fragments (HMF's). In this way we get a clear '*rise and fall*' in multifragmentation emission. The variation of the IMF's match with the experimental data when the distance between the nucleon is 4 fm for lower range of impact parameter. It is observed that IMF's shows the agreement with data at central collision but fails at semi & peripheral collision. This failure is due to the drawbacks in the method of analysis MST which we had used in our analysis at fixed momentum constraints. MST method gives the heavy cluster at the time of high densities. Further studies in this direction is in progress to optimize the value of $R_{clus}$ and momentum cut, so that better and fast method can be developed.

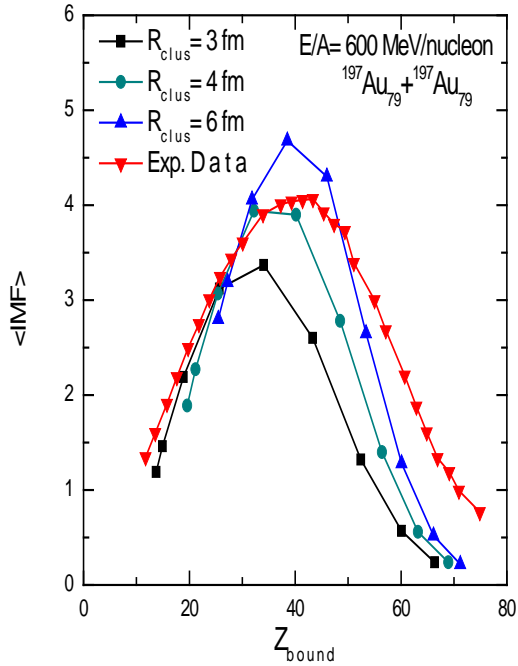

Fig.2.Multiplicity of IMF's as a function of $Z_{bound}$.

In Fig.2, comparison of theoretical calculations with experimental data of reactions $^{197}Au_{79}+^{197}Au_{79}$ at energy 600 MeV/nucleon[2] have been carried out. Here variation of IMF's as a function of $Z_{bound}$ have been displayed. One can see that for semi peripheral collisions multiplicity of <IMF> shows a peak because most of the spectator source does not take part in collision and large number of IMF's are